\documentclass[twocolumn,preprintnumbers,superscriptaddress,nofootinbib,aps,prd,floatfix]{revtex4-2}
\pdfoutput=1
\usepackage{enumerate}
\usepackage{amsmath,amssymb}
\usepackage{graphicx}
\usepackage{slashed}
\usepackage{float}
\usepackage[dvipsnames]{xcolor}
\usepackage[normalem]{ulem}
\usepackage{subfigure,orcidlink}
\usepackage{bbding}
\usepackage{multirow,array}
\usepackage{adjustbox}
\usepackage{hyperref}
\hypersetup{%
	colorlinks = true,%
	linkcolor = Blue,%
	citecolor = Blue,%
	filecolor = Blue,%
	urlcolor = Blue%
}
\usepackage{tabularray}
\UseTblrLibrary{booktabs}
\usepackage{tabularx}

\begin{document}
\title{Lepton parity dark matter and naturally unstable domain walls}
\author{Ernest Ma}
\affiliation{Department of Physics and Astronomy,
University of California, Riverside, California 92521, USA}

\author{Partha Kumar Paul$^{\orcidlink{https://orcid.org/0000-0002-9107-5635}}$}
\email{ph22resch11012@iith.ac.in}
\affiliation{Department of Physics, Indian Institute of Technology Hyderabad, Kandi, Telangana-502285, India.}

\author{Narendra Sahu$^{\orcidlink{https://orcid.org/0000-0002-9675-0484}}$}
\email{nsahu@phy.iith.ac.in}
\affiliation{Department of Physics, Indian Institute of Technology Hyderabad, Kandi, Telangana-502285, India.}

\date{\today}
 
\begin{abstract}
We propose a simple and predictive setup that connects neutrino masses, dark matter (DM), and gravitational waves. A minimal lepton parity DM scenario is considered where the residual symmetry $(-1)^L$ from the type I seesaw acts as the dark parity $D=(-1)^{L+2j}$, ensuring DM stability without imposing any new symmetry. A singlet Majorana fermion $S$ with even lepton parity serves as the DM candidate, interacting via a real scalar $\sigma$ which is also even lepton parity. The scalar potential possesses an accidental $\mathcal{Z}_2$ symmetry, whose spontaneous breaking gives rise to unstable domain walls (DW) in the presence of explicit $\mathcal{Z}_2$ breaking terms allowed by the lepton parity. The subsequent DW annihilation generates a stochastic gravitational wave (GW) background potentially observable at different GW experiments.
\end{abstract}
\maketitle
\noindent
\textit{Introduction.} In the standard model (SM) of quarks and leptons 
under the imposed $SU(3)_C \times SU(2)_L \times U(1)_Y$ gauge symmetry, it is well-known that the chosen particle content implies the 
automatic/accidental conservation of baryon number $B$ (under which 
quarks $\sim 1/3$) and lepton number $L$ (under which leptons $\sim 1$). 
If three singlet right-handed neutrinos (RHNs) are added with large Majorana masses, 
the well-known type I seesaw mechanism \cite{Minkowski:1977sc,Mohapatra:1980yp,Mohapatra:1991ng,Schechter:1980gr}  enables the observed left-handed 
neutrinos to be naturally light, and the associated conserved symmetry 
becomes $(-1)^L$, i.e. lepton parity.  As such, it can be used as dark 
symmetry, i.e. $D = (-1)^{L+2j}$, where $j$ is the spin of the particle 
concerned~\cite{Ma:2015xla}.

This insight~\cite{Ma:2015xla} allows a simple single additional particle to the SM 
as the dark matter (DM) of the Universe.  It may be a real scalar with odd 
$(-1)^L$ which has been studied in detail~\cite{GAMBIT:2017gge}, or a singlet 
Majorana fermion $S$ with even $(-1)^L$.  In the latter case, $S$ by itself 
decouples from the SM and requires another particle to make the connection. 
One choice is to use a scalar doublet with odd $(-1)^L$.  This becomes the 
well-known scotogenic model~\cite{Ma:2006km,Tao:1996vb} of radiative neutrino masses. 
Here we assume instead a real scalar $\sigma$ with even lepton parity. In this paper we show that $\sigma$ (which is necessary for $S$ to be 
DM) allows naturally unstable 
domain walls (DW), which are responsible for generating stochastic gravitational waves (GW) that can be observed at different GW experiments \cite{LISA:2017pwj,Garcia-Bellido:2021zgu,Sesana:2019vho,Adelberger:2005bt,Yunes:2008tw,LIGOScientific:2016wof,Weltman:2018zrl,Punturo:2010zz,NANOGrav:2023gor}.

In particle physics, discrete symmetries are imposed for various phenomenological purposes. If these discrete symmetries are broken spontaneously, then they can lead to stable DWs in the early Universe which can over close the Universe \cite{Zeldovich:1974uw,Kibble:1976sj,Vilenkin:1981zs,Kibble:1982dd,Lazarides:1981fv,Vilenkin:1984ib}. One way to avoid such a possibility is to add \textit{ad-hoc} explicit symmetry breaking terms (bias term) along with the spontaneous breaking of the discrete symmetry which can destabilize the DWs. We show that in a minimally extended type I seesaw, the residual lepton parity naturally allows such bias terms along with the spontaneous breaking term, leading to unstable DWs.\vspace{2mm}\\
\noindent
\textit{Minimal setup.} As discussed above, in the minimal type I seesaw model, the lepton number is explicitly broken to a residual symmetry $(-1)^L$ \footnote{In our model, the residual lepton parity can be identified as dark parity. This is shown in Appendix \ref{app:darkparity}.}.  By construction, the SM leptons are odd under this residual lepton parity. In this paper, we extend the minimal type I seesaw model with a right-handed fermion, $S$, with even lepton parity, thus it is disconnected from the SM. It naturally acts as a DM candidate. To produce it in the early Universe, we add a singlet scalar $\sigma$ with even lepton parity. Thus, it allows an interaction of kind $y_S\sigma\overline{S^c}S$ along with $y_R\sigma\overline{\nu^c_R}\nu_R$. The relevant Lagrangian of the model is
\begin{eqnarray}
    \mathcal{L}&=&\bar{S}i\gamma^\mu\partial_\mu S+\overline{\nu_R}i\gamma^\mu\partial_\mu\nu_R-\frac{y_S}{\sqrt{2}}\sigma \overline{S^c}S-\frac{y_R}{\sqrt{2}}\sigma \overline{\nu^c_R}\nu_R\nonumber\\&&-\frac{1}{2}M_R\overline{\nu^c_R}\nu_R-y_L\bar{L}\tilde{H}\nu_R-V(H,\sigma),\label{eq:lag}
\end{eqnarray}
where we have suppressed the generation indices of the RHNs.
In Eq \ref{eq:lag}, the most general scalar potential consisting of $H$ and $\sigma$ is given by $V(H,\sigma)=V_0 + V_1$ where
\begin{eqnarray}
    V_0&=&-\mu_H^2(H^\dagger H)+\lambda_H (H^\dagger H)^2-\frac{1}{2}\mu_\sigma^2\sigma^2+\frac{1}{4}\lambda_\sigma\sigma^4\nonumber\\&&+\frac{1}{2}\lambda_{H\sigma}(H^\dagger H)\sigma^2,
\end{eqnarray}
and
\begin{eqnarray}
    V_1=\mu_1\sigma^3/2\sqrt{2}+\mu_2H^\dagger H\sigma/\sqrt{2} .
\end{eqnarray}

Note that $V_0$ by itself possesses an accidental discrete $\mathcal{Z}_2$ symmetry under the transformation $\sigma\rightarrow-\sigma$, whereas $V_1$ breaks it softly by its linear and cubic terms, which may then be assumed naturally small. Note also that for positive $\mu_\sigma^2$ in $V_0$, there are two discrete nearly degenerate vacua, perfect for unstable DWs. If negative $\mu_\sigma^2$ is assumed \cite{Ma:2024dhk} so that $\sigma$ has a physical mass to begin with, as in most other studies of DM, this does not happen. We also note that the terms $\sigma\overline{S^c}S$ and $\sigma\overline{\nu_R^c}\nu_R$, break the $\mathcal{Z}_2$ symmetry explicitly while allowed by the dark parity. Therefore, we expect the corresponding Yukawa couplings $y_S$ and $y_R$ to be small. Since $y_S$ is small, the relic of $S$
can be produced through a non-thermal mechanism. We will discuss the DM relic in the next section.

When $\sigma$ acquires a vacuum expectation value (vev), the accidental $\mathcal{Z}_2$ symmetry breaks spontaneously and gives rise to DW networks, which are naturally unstable due to the presence of terms in $V_1$ yet allowed by lepton parity. In this framework, the light neutrino mass is given by
\begin{eqnarray}
    m_\nu=m_D(M_R+\sqrt{2}y_R v_\sigma)^{-1}m_D^T,
\end{eqnarray}
where $m_D=y_Lv_h/\sqrt{2}$, and $v_h$ is the SM Higgs vev.

In our setup, there are four free parameters, the vev of $\sigma$ ($v_\sigma$), mass of sigma ($M_\sigma$), $h-\sigma$ mixing ($\sin\theta$), and Yukawa coupling $y_{S}$, which play a role in the DM relic as well as in the GW analysis. While the small coupling $y_S$ and mixing $\sin\theta$ are responsible for the DM production via a non-thermal mechanism,  the $\sigma$ vev and $M_\sigma$ are responsible for the signature of the GW from the DW network.\vspace{2mm}\\
\noindent
\textit{Dark matter relic.}
The Majorana fermion $S$ acts as a natural dark matter (DM) candidate in our setup. The DM gains mass once the singlet scalar $\sigma$ obtains a vev after the spontaneous breaking of the accidental $\mathcal{Z}_2$ symmetry\footnote{We assume the bare mass of the DM to be negligible. This can be justified as its mass is generated by the $SS\sigma$ coupling and by $\sigma^3$ coupling in two loops. 
For the choice of model parameters, the loop contribution is very small even if the cutoff scale is the Planck mass. In our scenario $S$ is getting mass only via the vev of $\sigma$.}. The mass of the DM is given as
\begin{eqnarray}
M_S\equiv M_{\rm DM}=\sqrt{2}y_Sv_\sigma.\label{eq:mdm}
\end{eqnarray}
Note that the Yukawa coupling $y_S$ explicitly breaks the accidental $\mathcal{Z}_2$ symmetry and therefore must be small. This renders the DM relic abundance through the freeze-out mechanism difficult to achieve, as a small $y_S$ leads to an overabundant DM relic density and may even prevent thermalization. An alternative is to realize the correct relic abundance via non-thermal production from a heavier state, a process such as freeze-in \cite{McDonald:2001vt,Hall:2009bx,McDonald:2008ua} and SuperWIMP mechanism. Furthermore, the smallness of $y_S$ implies that the resulting DM mass is also light in this scenario. The singlet scalar $\sigma$  can decay into $S$ in our scenario. The SM Higgs can also produce $S$ through its mixing with  $\sigma$. Additionally, $S$ can be generated via $2\rightarrow2$ processes such as  $\sigma\sigma\rightarrow SS, hh\rightarrow SS, W^+W^-\rightarrow SS, ZZ\rightarrow SS, f\bar{f}\rightarrow SS$, where $f$ is the SM fermion. 
The DM relic density can be calculated by solving the coupled Boltzmann equations (BE)
\begin{eqnarray}
\frac{dY_S}{dz}&=&\frac{s(T)}{z\mathcal{H}(T)}\Bigg[\langle \sigma v\rangle_{{\rm SMSM}\rightarrow SS}\left((Y^{\rm eq}_{S})^2-Y_S^2\right)+\nonumber\\&& \langle \sigma v\rangle_{{\sigma\sigma}\rightarrow SS}\left(Y^{2}_{\sigma}-Y_S^2\frac{(Y^{\rm eq}_\sigma)^2}{(Y^{\rm eq}_S)^2}\right) +\nonumber\\&&2\frac{\langle\Gamma_\sigma\rangle}{s(T)} \left(Y_{\sigma}-Y_S\frac{Y^{\rm eq}_\sigma}{Y^{\rm eq}_S}\right) \Bigg],\label{eq:BES}
\end{eqnarray}
\begin{eqnarray}
\frac{dY_\sigma}{dz}&=&\frac{s(T)}{z\mathcal{H}(T)}\Bigg[-\langle \sigma v\rangle_{\sigma\sigma\rightarrow {\rm SMSM}}\left(Y_\sigma^2-(Y^{\rm eq}_{\sigma})^2\right)-\nonumber\\&& \langle \sigma v\rangle_{{\sigma\sigma}\rightarrow SS}\left(Y^{2}_{\sigma}-Y_S^2\frac{(Y^{\rm eq}_\sigma)^2}{(Y^{\rm eq}_S)^2}\right) -\nonumber\\&&2\frac{\langle\Gamma_\sigma\rangle}{s(T)} \left(Y_{\sigma}-Y_S\frac{Y^{\rm eq}_\sigma}{Y^{\rm eq}_S}\right) \Bigg],\label{eq:BEsigma}
\end{eqnarray}
where $z=M_\sigma/T$, $Y_i$ is the abundance of the $i^{\rm th}$ species defined as $Y_i=n_i/s$ where $n_i$ is the number density of that species and $s=\frac{2\pi^2}{45}g_{*s}T^3$ is the entropy density. $Y_i^{\rm eq}$ is the corresponding equilibrium abundance of that species. $\langle\Gamma_\sigma\rangle$ is the thermally averaged decay width of $\sigma$ to $S$\footnote{The SM Higgs also can decay to $S$. However, this decay mode is very small compared to $\sigma\rightarrow SS$; thus, we do not include it in our calculation.}. The factor of 2 accounts for the fact that each decay produces two DM particles. In the above equations, $\langle\sigma v\rangle_{ab\rightarrow cd}$ denotes the thermally averaged cross-section for a process $ ab\rightarrow cd$. $\mathcal{H}=1.66\sqrt{g_*}{T^2}/{M_{\rm pl}}$ is the Hubble parameter, $M_{\rm pl}=1.22\times10^{19}$ GeV is the Planck mass. $g_*$ is the relativistic degrees of freedom (d.o.f) at temperature $T$.
The DM relic density is then calculated to be
\begin{eqnarray}
\Omega_{\rm DM}h^2\simeq0.118\left(\frac{Y_S}{4.2\times10^{-10}}\right)\left(\frac{M_{\rm DM}}{1~\rm GeV}\right).
\end{eqnarray}
\begin{table*}[tbh]
\scalebox{1}{
\begin{tabular}{||c| c| c| c| c| c| c| c| c| c||} 
 \hline
  & $M_{\rm DM}$ (GeV) & $y_S$ & $v_\sigma$ (GeV) & $M_{\sigma}$ (GeV) &$\sin\theta$&$\sigma_{\rm DW}~(\rm TeV^3)$ &$T_{\rm ann}$ (GeV) & $\Omega_{\rm GW}^{\rm peak}h^2$ &$f_{\rm peak}$ (Hz) \\ [0.5ex] 
 \hline\hline
 BP1  & 0.148492 & $1.05\times10^{-9}$ & $10^{8}$ &$3\times10^7$ &$7.5\times10^{-6}$ &$2\times10^{14}$ & 120& $2.741\times10^{-7}$& $3.167\times10^{-6}$\\
 \hline
BP2  & 0.053 & $7.5\times10^{-9}$ & $5\times10^{6}$ &$8\times10^6$ &$3.8\times10^{-6}$ &$1.333\times10^{11}$ & 10& $3.232\times10^{-9}$& $2.559\times10^{-7}$\\
 \hline
BP3  & 1.061 & $7.5\times10^{-11}$ & $10^{10}$ &$5\times10^7$ &$7\times10^{-6}$ &$3.333\times10^{18}$ & $1.5\times10^4$& $3.007\times10^{-7}$& $3.976\times10^{-4}$\\
 \hline
BP4  & 7.071 & $10^{-10}$ & $5\times10^{10}$ &$10^8$ &$7\times10^{-6}$ &$1.667\times10^{20}$ & $1.1\times10^5$& $2.599\times10^{-7}$& $2.916\times10^{-3}$\\
 [1ex] 
 \hline
\end{tabular}}
\caption{Benchmark points giving rise to correct DM relic and observable GW signatures. $T_{\rm ann}$ is the annihilation temperature of the DW, $\sigma_{\rm DW}=\frac{2}{3}M_{\sigma}v_\sigma^2$ is the surface energy density of the DW, $\Omega_{\rm GW}^{\rm peak}h^2$ is the peak amplitude of GW, and $f_{\rm peak}$ is the peak frequency.}\label{tab:tab1}
\end{table*}
\begin{figure*}[!]
\centering
\includegraphics[scale=0.4]{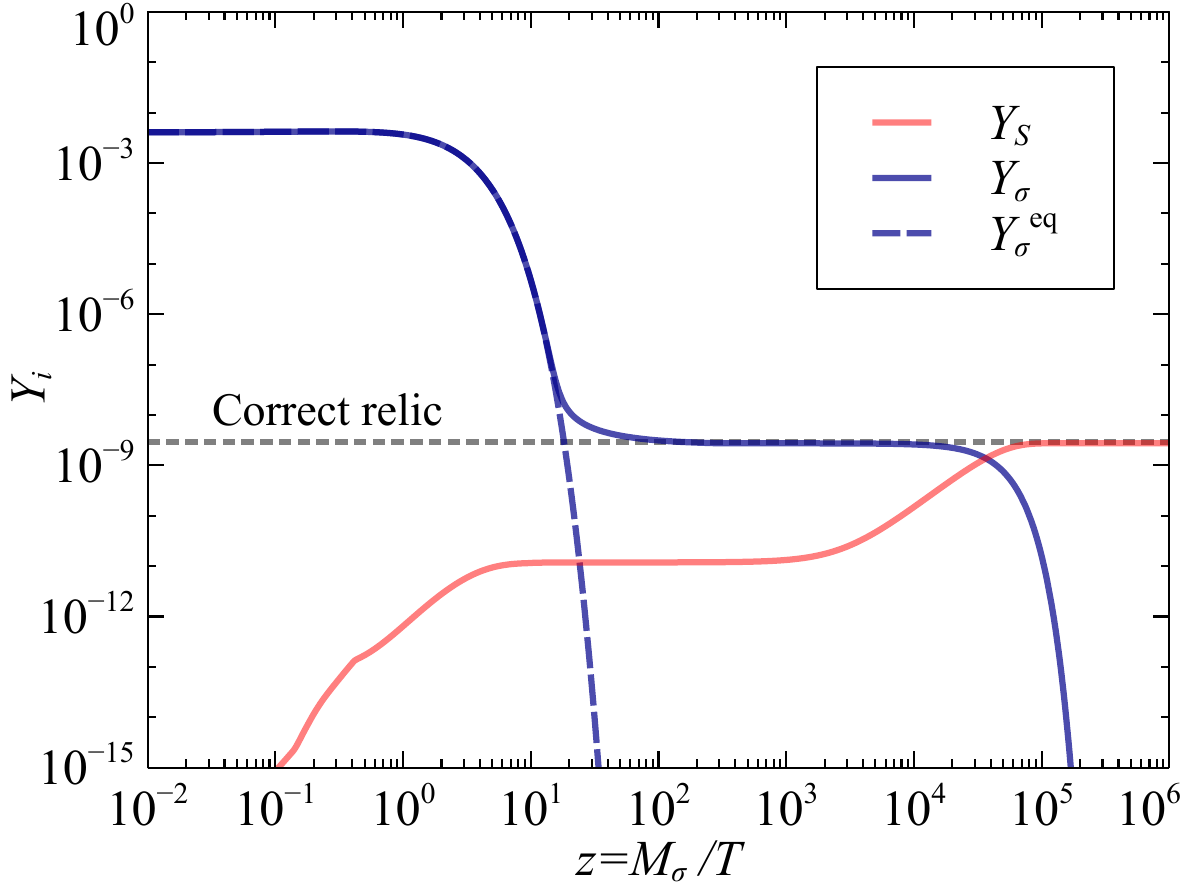}
\includegraphics[scale=0.4]{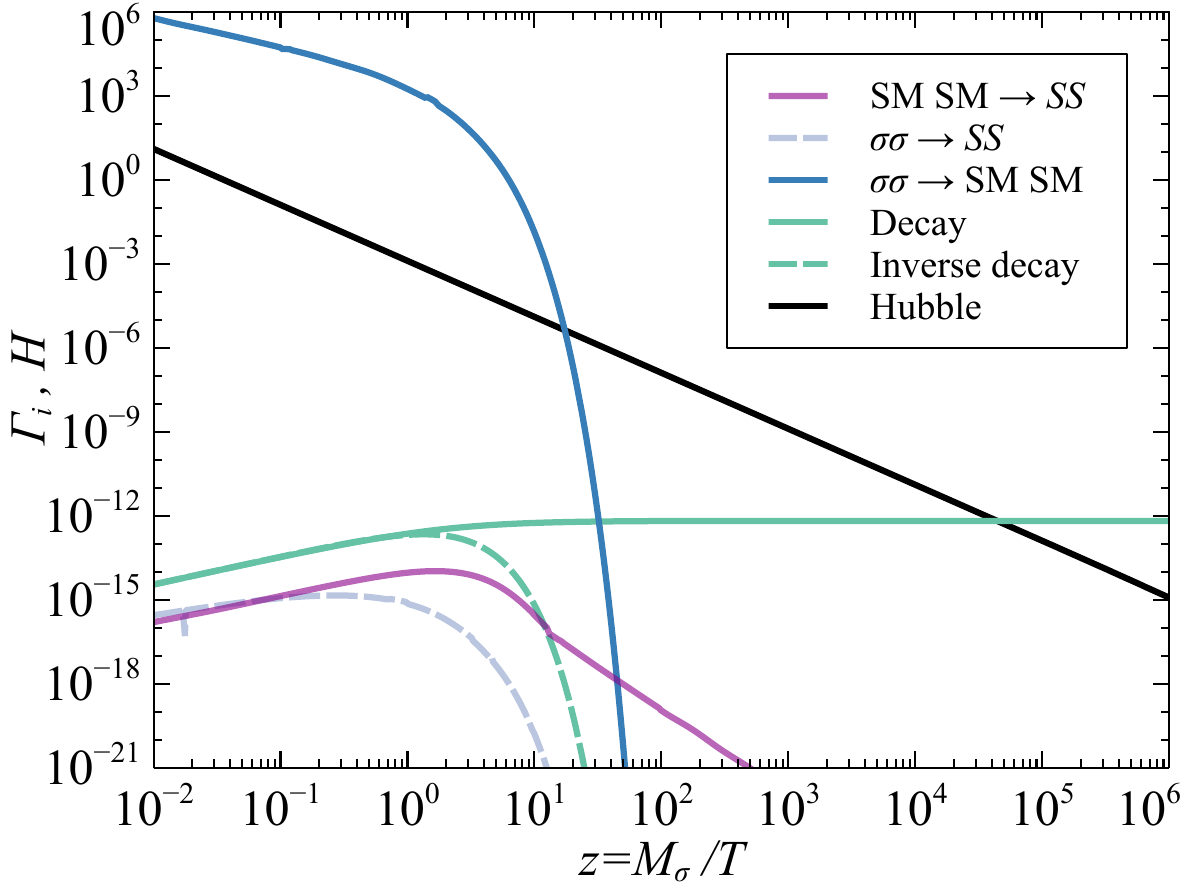}
\caption{\textit{Left:} Cosmological evolution of the DM ($S$) and the singlet scalar ($\sigma$) with respect to $z=M_\sigma/T$ for BP1 as mentioned in Table \ref{tab:tab1}. The blue dashed line represents the equilibrium abundance of $\sigma$, and the blue solid line represents the actual abundance of $\sigma$. The DM abundance is shown with the red solid line mentioned in the figure. The gray dashed line represents the correct relic of DM. \textit{Right:} Comparison of the interaction rates of different processes involved in the BE with the Hubble is shown with different colors as mentioned in the inset of the figure.}
\label{fig:evolu1}
\end{figure*}

As discussed earlier, DM can be produced both via the annihilation of SM particles and from the decay of $\sigma$. The scalar mixing angle suppresses the contribution from SM particle annihilation and remains subdominant unless the mixing angle is of suitable magnitude. DM can also be generated through $\sigma$ annihilations; however, the \textit{t}-channel contribution is suppressed by $y_S^4$, while the \textit{s}-channel is additionally suppressed by the $\sigma$ mass and the mixing angle. The singlet scalar $\sigma$ stays in thermal equilibrium with the SM bath through interactions such as ${\sigma\sigma \rightarrow W^+W^-, ZZ, hh,f\bar{f}}$. Once the rates of these processes drop below the Hubble expansion rate, $\sigma$ departs from equilibrium. The DM relic abundance is then predominantly built up from the subsequent out-of-equilibrium decay of $\sigma$. This corresponds to the SuperWIMP mechanism, where the $2\rightarrow2$ production channels are negligible compared to the late out-of-equilibrium decay contribution. Nevertheless, for certain parameter ranges with relatively large scalar mixing angles, the $2\rightarrow2$ production processes can become comparable to the out-of-equilibrium decay rate. In this regime, freeze-in and the SuperWIMP mechanism operate simultaneously, determining the final DM relic density.

In the \textit{left} panel of Fig. \ref{fig:evolu1}, we show the evolution of the DM abundance along with the $\sigma$ as a function of $z=M_\sigma/T$ for the BP1 as mentioned in Table \ref{tab:tab1}. We obtain the abundances by solving Eqs. \ref{eq:BES}, and \ref{eq:BEsigma} with an initial abundance of $S$ to be 0. The DM mass for this BP1 is 148.492 MeV. The interaction rates are compared with respect to Hubble in the \textit{right} panel of the same figure. Because of the small Yukawa coupling, the DM never reaches equilibrium. This can be easily understood from the \textit{right} panel of Fig. \ref{fig:evolu1} as the ${\rm SM~SM}\rightarrow SS$ rate remains much below the Hubble rate shown with the magenta solid line.  The interaction rate of $\sigma\sigma\rightarrow{\rm SM~SM}$ remains above the Hubble up to $z\sim20$. Thus, $\sigma$ goes out of equilibrium at that time and its relic gets frozen. The decay rate goes above the Hubble around $z\sim4\times10^4$. As a result, the DM gets maximally produced from this decay around this time. Before the out-of-equilibrium decay of the $\sigma$, DM also gets produced from the $2\rightarrow2$ processes. This $2\rightarrow2$ production is minimal compared to the decay of $\sigma$. The DM relic gets frozen at around $z\sim8\times10^4$, corresponding to a temperature of $T\sim375$ GeV. This temperature is much above the mass scale of the DM.

The model evades current direct detection bounds due to the extremely small Yukawa coupling. In this scenario, a typical annihilation cross-section into various SM states (\textit{i.e,} $SS\rightarrow W^+W^-,e^+e^-$, etc.) is likewise highly suppressed, ensuring consistency with existing indirect detection limits from CMB \cite{Madhavacheril:2013cna,Slatyer:2015jla}. Nevertheless, the DM mass is intrinsically linked to the properties of the DWs, making the scenario testable through GW observations, as discussed in the latter part of the paper.
\begin{figure*}[!]
\centering
\includegraphics[scale=0.42]{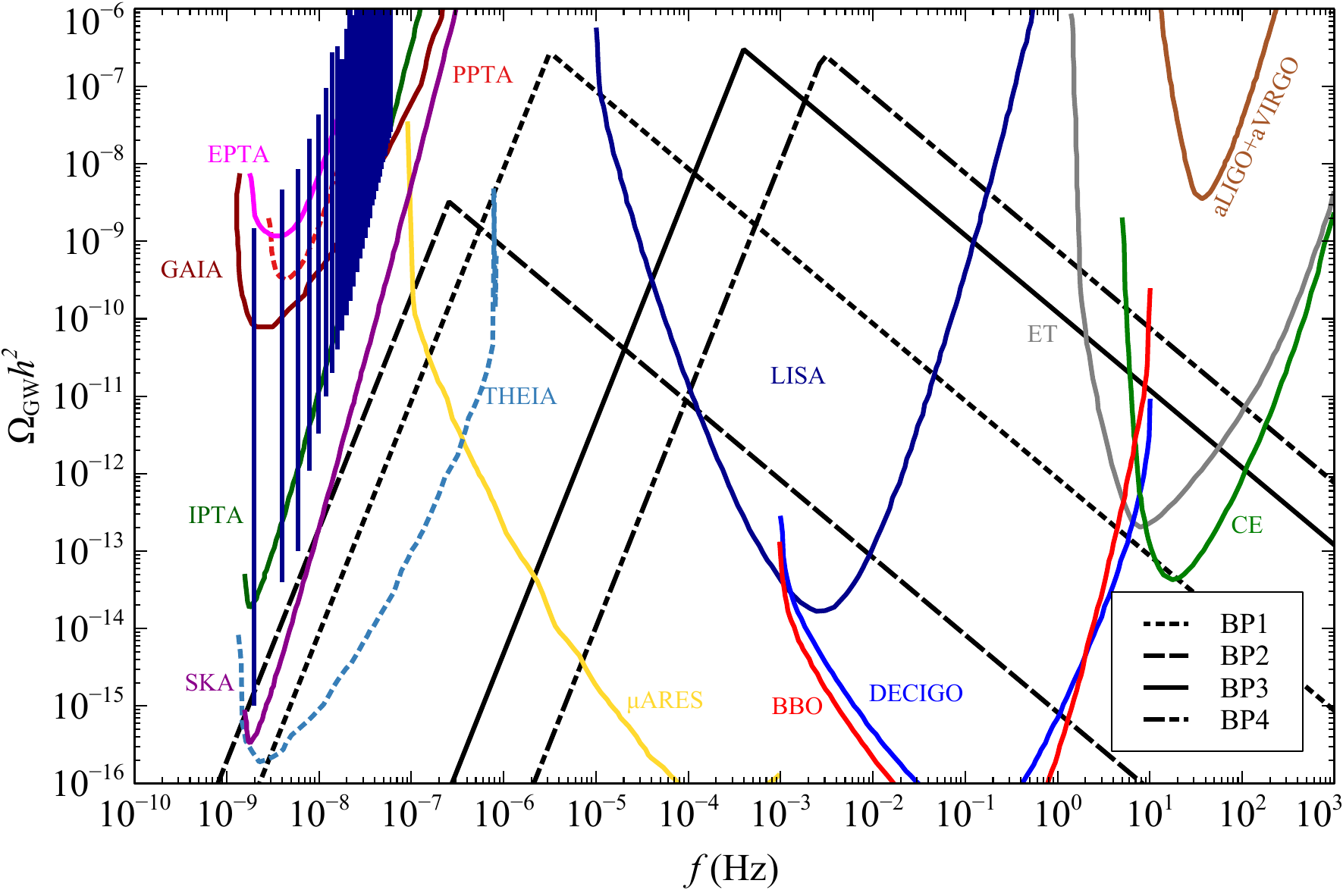}
\caption{GW spectrum is shown for BP1, BP2, BP3 and BP4 as mentioned in Table \ref{tab:tab1}. The sensitivities from different GW experiments are shown with different colored lines.}
\label{fig:gwbp}
\end{figure*}
\vspace{2mm}\\
\noindent
\textit{Gravitational waves from domain walls.}
The accidental $\mathcal{Z}_2$ symmetry can be spontaneously broken by the vev of $\sigma$. The spontaneous breaking of the discrete symmetry leads to the formation of domain walls in the early Universe, which are two-dimensional topological defects in spacetime. These DWs overclose the energy density of the Universe very soon, as their energy density scales in the scaling regime as $a^{-2}$, while the matter and radiation scales as $a^{-3}$ and $a^{-4}$, respectively, where $a$ is the scale factor of the Universe. DWs can be made unstable if a pressure difference is created between the two vacua of the potential. The presence of small explicit $\mathcal{Z}_2$ breaking terms in $V_1$ acts as a bias potential that lifts the minima by
\begin{eqnarray}
    V_{\rm bias}=\frac{\mu_1v_\sigma^3}{\sqrt{2}}+\frac{\mu_2v_\sigma v_h^2}{\sqrt{2}}.
\end{eqnarray}
This causes the DWs to collapse and release their energy in the form of GWs. The DW has to disappear before the Big Bang Nucleosynthesis (BBN) and before it can dominate the energy density of the Universe. The details are given in the Appendix \ref{app:dwgw}. The DW annihilation dynamics and subsequent release of GW depend on two crucial parameters: the annihilation temperature, $T_{\rm ann}$ at which the DW annihilate, and the surface energy density or surface tension of the DW, $\sigma_{\rm DW}=\frac{2}{3}M_{\sigma}v_\sigma^2$. The larger the surface tension, the larger the GW amplitude. On the other hand, if the DW annihilates at an early epoch \textit{i.e.} at a larger temperature, then the amplitude will be smaller.

We then compute the GW spectrum for the benchmark points listed in Table \ref{tab:tab1} and displayed in Fig. \ref{fig:gwbp}. For BP1, the $\sigma$ vev and mass are $10^8$ GeV and $3\times10^7$ GeV, respectively, which yield a surface tension of $\sigma_{\rm DW}=2\times10^{14}~\rm TeV^3$. The DW domination criterion imposes a lower bound of 116 GeV on the annihilation temperature. Hence, we select 120 GeV as the annihilation temperature for this BP1\footnote{We can obtain the desired $T_{\rm ann}$ by appropriately varying $\mu_1$ and $\mu_2$.}. This choice gives a peak frequency of $3.167\times10^{-6}$ Hz and a peak GW amplitude of $2.741\times10^{-7}$. BP1 lies within the sensitivity ranges of LISA \cite{LISA:2017pwj}, THEIA \cite{Garcia-Bellido:2021zgu}, $\mu$ARES \cite{Sesana:2019vho}, DECIGO \cite{Adelberger:2005bt}, BBO \cite{Yunes:2008tw}, and CE \cite{LIGOScientific:2016wof}. For BP2, the surface tension is $1.333\times10^{11}~{\rm TeV^3}$, which sets a lower bound of 3.2 GeV on the annihilation temperature. We take the annihilation temperature to be 10 GeV, resulting in a peak frequency of $2.559\times10^{-7}$ Hz and a peak GW amplitude of $3.232\times10^{-9}$. BP2 falls within the sensitivity bands of SKA \cite{Weltman:2018zrl}, THEIA, $\mu$ARES, DECIGO, and BBO. The tail of its spectrum can also accommodate the NANOGrav \cite{NANOGrav:2023gor} data shown by the blue vertical lines. For BP3, the lower limit on the annihilation temperature is 14912 GeV. We choose $1.5\times10^4$ GeV as the annihilation temperature and compute the GW spectrum. This BP3 lies in the sensitivity ranges of LISA, $\mu$ARES, DECIGO, BBO, CE, and ET \cite{Punturo:2010zz}. BP4 corresponds to $v_\sigma=5\times10^{10}$ GeV and $M_\sigma=10^8$ GeV, giving a surface energy density of $1.667\times10^{20}~\rm TeV^3$. The annihilation temperature must exceed 105444 GeV, and we consider it to be $1.1\times10^5$ GeV. BP4 lies within the sensitivity ranges of $\mu$ARES, LISA, DECIGO, BBO, CE, and ET.

If we assume DW annihilation takes place just above the wall domination epoch, \textit{i.e.} $T_{\rm dom}\lesssim T_{\rm ann}$, which corresponds to the lowest achievable annihilation temperature of the DW, then we observe a correlation between the DM mass and the lowest peak frequency of the GW \textit{i.e.}  the peak frequency of the GW decreases as the DM mass decreases and vice versa. This can be understood as follows. From Eq. \ref{eq:mdm}, we see that a decrease in DM mass corresponds to a lowering in $\sigma$ vev. A smaller $\sigma$ vev delays the wall domination epoch, as $t_{\rm dom}\propto\frac{1}{\sigma_{\rm DW}}\sim\frac{1}{M_\sigma v_\sigma^2}$, leading to a lower wall domination temperature. Consequently, the annihilation temperature drops as the vev decreases, and the frequency decreases accordingly since $f_{\rm peak}\propto T_{\rm ann}$.  To demonstrate the correlation between the DM mass and GW peak frequency, let us consider three DM masses such that $M^{(1)}_{\rm DM}<M^{(2)}_{\rm DM}<M^{(3)}_{\rm DM}$. This leads to $T^{(1)}_{\rm dom}<T^{(2)}_{\rm dom}<T^{(3)}_{\rm dom}$. Since we assume that $T_{\rm ann}$ is very close to $T_{\rm dom}$ and  follows a similar trend as $T_{\rm dom}$, \textit{i.e.} $T^{(1)}_{\rm ann}<T^{(2)}_{\rm ann}<T^{(3)}_{\rm ann}$, then $f^{(1)}_{\rm peak}<f^{(2)}_{\rm peak}<f^{(3)}_{\rm peak}$. This can be easily checked by comparing BP1, BP2, BP3, and BP4 in Table \ref{tab:tab1}. We note that, in our case, the DM mass is related to the lowest peak frequency of the GW due to our assumption that $T_{\rm ann}\gtrsim T_{\rm dom}$. However, we note that $T_{\rm ann}$ is independent of DM mass and in general can be very large ($T_{\rm ann}\gg T_{\rm dom}$) and may not follow the above-mentioned trend. In such a case, there is no correlation between the DM mass and GW peak frequency.\vspace{2mm}\\
\noindent
\textit{Conclusions.} We have proposed a minimal and predictive framework connecting neutrino masses, dark matter (DM), and gravitational waves. The residual lepton parity $(-1)^L$ of the type I seesaw is identified with the dark parity $D=(-1)^{L+2j}$, ensuring DM stability without introducing any additional symmetry. In this setup, a singlet Majorana fermion $S$ acts as the DM candidate and interacts through a real scalar $\sigma$. The scalar potential involving $\sigma$ exhibits an accidental $\mathcal{Z}_2$ symmetry whose spontaneous breaking leads to the formation of domain walls. Explicit $\mathcal{Z}_2$ breaking terms allowed by lepton parity create a pressure difference across the walls, making them annihilate, generating a stochastic GW background within the sensitivity of upcoming GW experiments. This is an important result of this paper in comparison to the existing literature where \textit{ad-hoc} explicit symmetry breaking terms are used to destabilize the DW \cite{Vilenkin:2000jqa,Gelmini:1988sf,Larsson:1996sp,Saikawa:2017hiv,Nakayama:2016gxi,Paul:2024iie,Hiramatsu:2013qaa}.

The model naturally harbors a light DM ($S$) whose relic can be established via freeze-in and/or SuperWIMP mechanism. The DM mass arises solely from the Yukawa term $y_S S S \sigma$ once $\sigma$ acquires a vacuum expectation value. Since this operator softly breaks the accidental $\mathcal{Z}_2$, the coupling $y_S$ must be extremely small. This small $y_S$ naturally suppresses the DM mass, rendering it light. The same small coupling also makes the standard freeze-out mechanism ineffective, pointing to non-thermal production as the viable mechanism to obtain the observed relic density. 

Another intriguing outcome of this framework is that there can be a correlation between the DM mass and the peak frequency of the GW spectrum if we assume that $T_{\rm ann}\gtrsim T_{\rm dom}$, where $T_{\rm ann}$ is related to the peak frequency of the GW and $T_{\rm dom}$ is related to DM mass in our scenario. Note that this correlation is limited to the lowest peak frequency of the GW spectrum. However, in principle $T_{\rm ann}$ is independent of DM mass and the annihilation of the DW can happen at any temperature, say $T_{\rm ann}\gg T_{\rm dom}$ which corresponds to a larger peak frequency of the GW. In that case, there is no correlation between the peak frequency of GW and the DM mass. Although the precise annihilation temperature depends on additional model parameters, this qualitative trend establishes a testable link between particle physics and cosmology, where future GW observations could indirectly probe the light DM in this scenario.
\vspace{2mm}\\
\noindent
\textit{Acknowledgments.} P.K.P. would like to acknowledge the Ministry of Education, Government of India, for providing financial support for his research via the Prime Minister’s Research Fellowship (PMRF) scheme. 

\appendix
\section{LEPTON PARITY $(-1)^L$ IDENTIFIED AS DARK PARITY, $D=(-1)^{L+2j}$}\label{app:darkparity}
The charges of particles under $U(1)_L$, $(-1)^L$, and $(-1)^{L+2j}$ are given in Table \ref{tab:tab2}. The allowed terms by these symmetries are given in Table \ref{tab:tab3}.
\begin{table}[h]
\centering\caption{}
\scalebox{0.8}{
\begin{tabular}{||c| c| c| c| c| c||} 
 \hline
 Symmetry & $S$ & $\sigma$ & $\nu_R$ & $L$ &$H$ \\ [0.5ex] 
 \hline\hline
$L$  & 0 & 0 & 1 &1 &0\\[0.7ex] 
 \hline
 $(-1)^L$  & + & + & -- &-- &+\\[0.7ex] 
 \hline
 $D=(-1)^{L+2j}$  & -- & + & + &+ &+\\[0.7ex] 
 \hline
\end{tabular}}
\label{tab:tab2}
\end{table}
\begin{table}[h]
\centering\caption{}
\scalebox{0.8}{
\begin{tabular}{||c| c| c| c||} 
 \hline
 Terms & $L$ & $(-1)^L$ & $D=(-1)^{L+2j}$\\ [0.8ex] 
 \hline\hline
$\overline{S^c}S\sigma$  &\textcolor{blue}{\CheckmarkBold}  & \textcolor{blue}{\CheckmarkBold}   & \textcolor{blue}{\CheckmarkBold}  \\[0.8ex] 
 \hline
$\overline{\nu_R^c}\nu_R\sigma$  &  \textcolor{red}{\XSolidBrush}  & \textcolor{blue}{\CheckmarkBold}  & \textcolor{blue}{\CheckmarkBold}  \\[0.7ex] 
 \hline
 $\overline{S}\nu_R\sigma$  &  \textcolor{red}{\XSolidBrush}  &  \textcolor{red}{\XSolidBrush}  &  \textcolor{red}{\XSolidBrush} \\[0.7ex] 
 \hline
 $\bar{L}\tilde{H}\nu_R$  & \textcolor{blue}{\CheckmarkBold} & \textcolor{blue}{\CheckmarkBold}  & \textcolor{blue}{\CheckmarkBold} \\[0.7ex] 
 \hline
 $\bar{L}\tilde{H}S$  &  \textcolor{red}{\XSolidBrush}  &  \textcolor{red}{\XSolidBrush}  &  \textcolor{red}{\XSolidBrush} \\[0.7ex] 
 \hline
\end{tabular}
\label{tab:tab3}}
\end{table}

From Table \ref{tab:tab2}  and \ref{tab:tab3}, it is clear that the residual lepton parity, $(-1)^L$ acts as dark parity, $D=(-1)^{L+2j}$ in our model.

\section{SCALAR MIXING}
The scalar potential involving both the scalars can be written as
\begin{eqnarray}
   V(H,\sigma)&=&-\mu_H^2(H^\dagger H)+\lambda_H (H^\dagger H)^2-\frac{1}{2}\mu_\sigma^2\sigma^2+\frac{1}{4}\lambda_\sigma\sigma^4\nonumber\\&&+\frac{1}{2}\lambda_{H\sigma}(H^\dagger H)\sigma^2.  
\end{eqnarray}
The scalars can be parametrized as
\begin{eqnarray}
    H=\begin{pmatrix}
        0\\ \frac{h+v_h}{\sqrt{2}}
    \end{pmatrix},~\sigma=\sigma+v_\sigma
\end{eqnarray}
The mass-squared matrix is obtained to be
\begin{eqnarray}
    \mathcal{M}^2=\begin{pmatrix}
        2v_h^2\lambda_H& v_hv_\sigma\lambda_{H\sigma}\\
        v_hv_\sigma\lambda_{H\sigma}& 2v_\sigma^2\lambda_\sigma
    \end{pmatrix},
\end{eqnarray}
which can be diagonalized with the following matrix
\begin{eqnarray}
    R=\begin{pmatrix}
        \cos\theta&-\sin\theta\\
        \sin\theta&\cos\theta
    \end{pmatrix},
\end{eqnarray}
where the mixing angle is given as
\begin{eqnarray}
\tan2\theta=\frac{v_hv_\sigma\lambda_{H\sigma}}{v_\sigma^2\lambda_\sigma-v_h^2\lambda_h}.
\end{eqnarray}
This leads to two mass states $h_1$ and $h_2$ with masses $M_{h_1}$, and $M_{h_2}$, where $h_1$ is identified to be the SM Higgs and $h_2$ is the $\sigma$.
We can express the scalar couplings in terms of the physical scalar masses and mixing angle as
\begin{eqnarray}
    \lambda_H=\frac{M_{h_1}^2\cos^2\theta+M_{h_2}^2\sin^2\theta}{2v_h^2},
\end{eqnarray}
\begin{eqnarray}
    \lambda_\sigma=\frac{M_{h_2}^2\cos^2\theta+M_{h_1}^2\sin^2\theta}{2v_\sigma^2},
\end{eqnarray}
\begin{eqnarray}
    \lambda_{H\sigma}=\sin\theta\cos\theta\frac{M^2_{h_2}-M^2_{h_1}}{v_hv_\sigma}.
\end{eqnarray}
Since $\sin\theta$ is small, for all practical purposes, we use the notation $M_{h_2}\equiv M_\sigma$.

\section{DOMAIN WALL FROM $\mathcal{Z}_2$ SYMMETRY BREAKING AND PRODUCTION OF GRAVITATIONAL WAVES}\label{app:dwgw}
The scalar fields are parametrized as,
\begin{eqnarray}
H=\begin{pmatrix}
    0\\ \frac{h+v_h}{\sqrt{2}}
\end{pmatrix}, ~\sigma=\sigma+v_\sigma.
\end{eqnarray}

Then the scalar potential can be written as
\begin{eqnarray}
	V(h,\sigma)&=&-\frac{\mu^2_H}{2}h^2+\frac{\lambda_H}{4}h^4-\frac{\mu^2_\sigma}{2}\sigma^2+\frac{\lambda_\sigma}{4}\sigma^4\nonumber\\&&+\frac{\lambda_{H\sigma}}{4}h^2\sigma^2\label{eq:rhopot}
\end{eqnarray}

The equations of motion for the DW are
\cite{Vilenkin:2000jqa,Gelmini:1988sf,Larsson:1996sp,Saikawa:2017hiv,Nakayama:2016gxi,Paul:2024iie},
\begin{eqnarray}
\frac{d^2h}{dx^2}-\frac{d V}{d h}=0,~\frac{d^2\sigma}{dx^2}-\frac{d V}{d \sigma}=0,
\end{eqnarray}
with the boundary condition
\begin{eqnarray}
\lim_{x\rightarrow\pm\infty}h(x)= v_h,~~\lim_{x\rightarrow\pm\infty}\sigma(x)=\pm v_\sigma.
\end{eqnarray}

In the decoupling limit, after solving the equation of motion, we obtain,
\begin{eqnarray}
\sigma(x)=v_\sigma\tanh(\alpha x),
\end{eqnarray}
where $\alpha\simeq\sqrt{\frac{\lambda_\sigma}{2}}v_\sigma$.

The DW is extended along the $x=0$ plane, and the two vacua are realized at $x\rightarrow\pm\infty$. The width of the DW is estimated as $\delta\sim\left(\frac{\sqrt{\lambda_\sigma}v_\sigma}{\sqrt{2}}\right)^{-1}$. The surface energy density, also referred to as the tension of the DWs, is calculated to be,
\begin{eqnarray}
\sigma_{\rm DW}=\frac{4}{3}\sqrt{\frac{\lambda_\sigma}{2}}v_\sigma^3\simeq\frac{2}{3}M_{\sigma}v_{\sigma}^2,
\end{eqnarray}
where $M_{\sigma}=\sqrt{2\lambda_\sigma}v_\sigma$.
As discussed earlier, without a soft $Z_2$ breaking term, the DW will be stable and will overclose the energy density of the Universe.
To overcome this problem, we introduce an energy bias in the potential as $\mu_{1}\sigma^3/2\sqrt{2}+\mu_{2}\sigma h^2/2\sqrt{2}$, which breaks the $Z_2$ symmetry explicitly. Here $\mu_{1},\mu_2$ are mass dimension one couplings. The Eq \ref{eq:rhopot} then becomes
\begin{eqnarray}
\mathcal{V}=V(h,\sigma)+\mu_{1}\sigma^3/2\sqrt{2}+\mu_{2}\sigma h^2/2\sqrt{2}.
\end{eqnarray}
 As a result, the degeneracy of the minima is lifted by
\begin{eqnarray}
V_{\rm bias}\equiv |\mathcal{V}(-v_\sigma)-\mathcal{V}(v_\sigma)|=\frac{\mu_1v_\sigma^3}{\sqrt{2}}+\frac{\mu_2v_\sigma v_h^2}{\sqrt{2}}.
\end{eqnarray}
The energy bias has to be large enough so that the DW annihilate before the BBN epoch and before they can dominate the energy density of the Universe. The annihilation time scale of the DW is given by
\begin{eqnarray}
    t_{\rm ann}=\mathcal{C}_{\rm ann}\frac{\mathcal{A}\sigma_{\rm DW}}{V_{\rm bias}},\label{eq:tann}
\end{eqnarray}
where $\mathcal{C}_{ann}$ is a coefficient of $\mathcal{O}(1)$, $\mathcal{A}\simeq0.8\pm0.1$\cite{Hiramatsu:2013qaa} is area parameter. The timescale at which the DW will dominate the energy density of the Universe is given as
\begin{eqnarray}
    t_{\rm dom}=\frac{3M_{\rm pl}^2}{32\pi \mathcal{A}\sigma_{\rm DW}}\label{eq:tdom}
\end{eqnarray}
The DW must annihilate before they dominate the energy density of the Universe, which is set by
\begin{eqnarray}
    t_{\rm ann}<t_{\rm dom}.\label{eq:tanndom}
\end{eqnarray}
Assuming the annihilation happens during the radiation-dominated era, we have
\begin{eqnarray}
    t_{\rm ann}=\frac{1}{2\mathcal{H}(T_{\rm ann})}=\frac{M_{\rm pl}}{2\times1.66\sqrt{g_*}T^2_{\rm ann}}\label{eq:Tanntann}
\end{eqnarray}
This results in a lower limit on the annihilation temperature to be
\begin{eqnarray}
    T_{\rm ann}>3.18g_{*}^{-1/4}M_{\rm pl}^{-1/2}\mathcal{A}^{1/2}\sigma_{\rm DW}^{1/2}
\end{eqnarray}

The wall domination criteria give a lower bound on the bias potential to be
\begin{eqnarray}
    V_{\rm bias}>\frac{32\pi}{3}\mathcal{C}_{\rm ann}\frac{\mathcal{A}^2\sigma_{\rm DW}^2}{M_{\rm pl}^2}.
\end{eqnarray}

This puts a lower limit on the bias parameters $\mu_1$ and $\mu_2$ as
\begin{eqnarray}
    \mu_1 v_\sigma^2+\mu_2 v_h^2>67.4\frac{M_\sigma^2 v_\sigma^3}{M_{\rm pl}^2}.
\end{eqnarray}

The DWs must also annihilate before the BBN to be consistent with the BBN prediction. This puts a lower limit on the bias potential to be
\begin{eqnarray}
    V_{\rm bias}>\frac{\mathcal{C}_{\rm ann}\mathcal{A}\sigma_{\rm DW}}{\tau_{\rm BBN}},
\end{eqnarray}
where $\tau_{\rm BBN}$ is the BBN time scale. Consequently, the bias parameters $\mu_1$ and $\mu_2$ are constrained from above as
\begin{eqnarray}
    \mu_1 v_\sigma^2+\mu_2 v_h^2>3.77\frac{M_\sigma v_\sigma}{\tau_{\rm BBN}}.
\end{eqnarray}

It has to be noted that the bias potential $V_{\rm bias}$ can not be arbitrarily large due to the requirement of percolation of
both the vacua \cite{Gelmini:1988sf}. This gives an upper bound on the bias potential to be
\begin{eqnarray}
    V_{\rm bias}<0.795V_0,
\end{eqnarray}
where $V_0$ is the potential difference between the maximum and the $+v_\sigma$ minimum of the potential. This translates into an upper limit on the bias parameters $\mu_1$ and $\mu_2$ as
\begin{eqnarray}
    \mu_1 v_\sigma^2+\mu_2 v_h^2<0.193\lambda_\sigma v_\sigma^3.
\end{eqnarray}
\begin{figure}[H]
    \centering
    \includegraphics[scale=0.45]{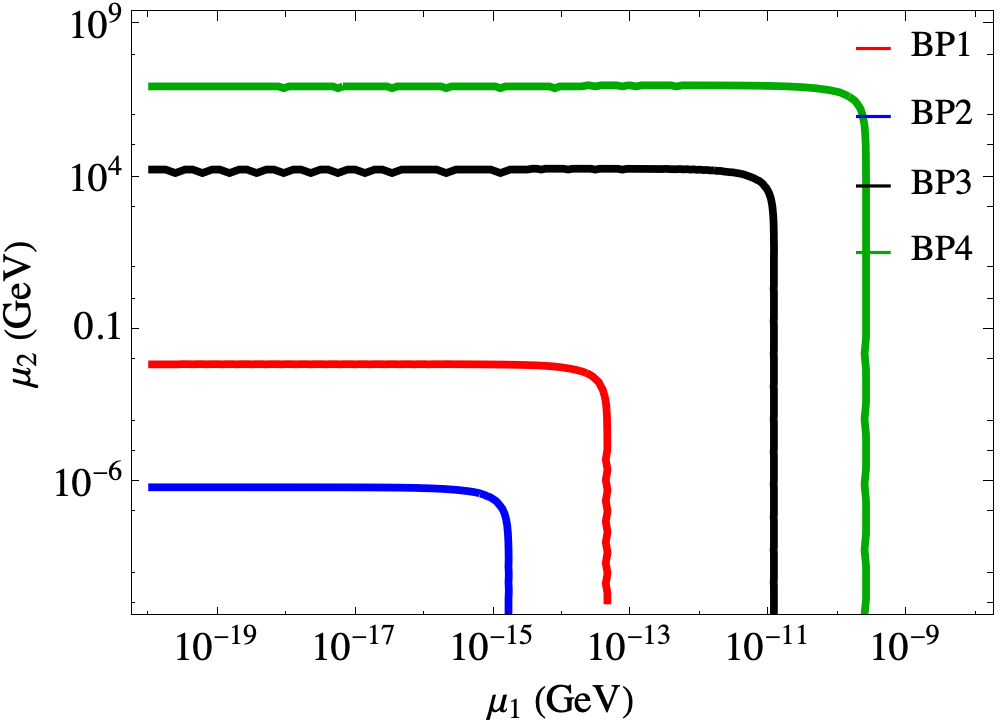}
    \caption{Allowed values of $\mu_1$ and $\mu_2$ which give the chosen $T_{\rm ann}$ as mentioned in Table \ref{tab:tab1} for four BPs.}
    \label{fig:mu1vsmu2}
\end{figure}

In Fig \ref{fig:mu1vsmu2}, we show the allowed values of the bias parameters which give the DW annihilation temperatures, $T_{\rm ann}$ as mentioned in the Table \ref{tab:tab1}. The four contours represents the four BPs.

The peak amplitude of the GW spectrum at present, $t_0$, is given by \cite{Saikawa:2017hiv}
\begin{eqnarray}
	\Omega^{\rm peak}_{\rm GW}h^2(t_0)&=&7.18824\times10^{-18}\mathcal{A}^2\tilde{\epsilon}_{GW} \bigg(\frac{\sigma_{\rm DW}}{1{\rm TeV^3}}\bigg)^2\nonumber\\&&\bigg(\frac{g_{*s}(T_{\rm ann})}{10}\bigg)^{-\frac{4}{3}}\bigg(\frac{T_{\rm ann}}{10^{-2}\rm GeV}\bigg)^{-4},\label{eq:peakamp}
\end{eqnarray}
where $\tilde{\epsilon}_{GW}\simeq0.7\pm0.4$\cite{Hiramatsu:2013qaa} is the efficiency parameter, $T_{\rm ann}$ is the temperature at which the DWs annihilate, $g_{*s}(T_{\rm ann})$ is the relativistic entropy d.o.f at the epoch of DWs annihilation.

Assuming the DWs disappear at temperature $T_{\rm ann}$, the peak frequency of the GW spectrum at present is estimated as 
\begin{eqnarray}
	f_{\rm peak}(t_0)&=&1.78648\times10^{-10}{\rm Hz}\bigg(\frac{g_{*s}(T_{\rm ann})}{10}  \bigg)^{-\frac{1}{3}}\nonumber\\&&\bigg(\frac{g_{*}(T_{\rm ann})}{10}\bigg)^{\frac{1}{2}}\bigg(\frac{T_{\rm ann}}{10^{-2}\rm GeV}\bigg).\label{eq:peakfre}
\end{eqnarray}
The amplitude of the GW for any frequency at present varies as
\begin{equation}
\Omega_{\rm GW}h^2(t_0,f) =\Omega^{\rm peak}_{\rm GW}h^2(t_0)\left\{
	\begin{array}{l}
	\frac{f_{\rm peak}}{f}~~~~~~,~f>f_{\rm peak}\\
	\bigg(\frac{f}{f_{\rm peak}}\bigg)^3,~f<f_{\rm peak}.\\
	\end{array}
	\right.
\end{equation}

\section{THERMALLY AVERAGED CROSS-SECTION}
The thermally averaged cross-section for a process, $ab\rightarrow cd$ is given by
\begin{eqnarray}
\langle\sigma v \rangle_{ab\rightarrow cd}&=&\frac{T}{32\pi^4}\frac{g_ag_b}{n_a^{\rm eq}n_{b}^{\rm eq}}\int_{(m_a+m_b)^2}^{\infty}\sigma_{ab\rightarrow cd}(s)\sqrt{s} K_1\left(\frac{\sqrt{s}}{T}\right) \nonumber\\&&\times\frac{\left(s-(m_a+m_b)^2\right)\left(s-(m_a-m_b)^2\right)}{4s}ds,
\end{eqnarray}
where $n_a^{\rm eq},n_b^{\rm eq}$ are the equilibrium number densities of the $a$ and $b$ particles, $m_a,m_b$ are the mass of the respective particles, and $g_a,g_b$ are the d.o.f of the particles. $s$ is the center of mass energy, and $K_1$ is the modified Bessel function of the first kind. We used {\tt CalcHEP} \cite{Belyaev:2012qa} for calculating the cross-sections of the relevant processes involved in DM relic calculations.

\section{DECAY WIDTH OF THE SCALAR}
The decay width of $\sigma$ to the DM, $S$ is calculated to be
\begin{eqnarray}
    \Gamma_\sigma=\cos^2\theta\frac{y_S^2}{16\pi}M_\sigma\left(1-\frac{4M_S^2}{M_\sigma^2}\right)^{3/2}.
\end{eqnarray}
The thermally averaged decay rate is given as
\begin{eqnarray} \langle\Gamma_\sigma\rangle=\Gamma_\sigma\frac{K_1\left(\frac{M_\sigma}{T}\right)}{K_2\left(\frac{M_\sigma}{T}\right)},
\end{eqnarray}
where $K_2$ is the modified Bessel function of the second kind.

\section{EVOLUTION OF DARK MATTER RELIC AND INTERACTION RATES FOR BP2, BP3, AND BP4}
\begin{figure*}[h]
\centering
\includegraphics[scale=0.4]{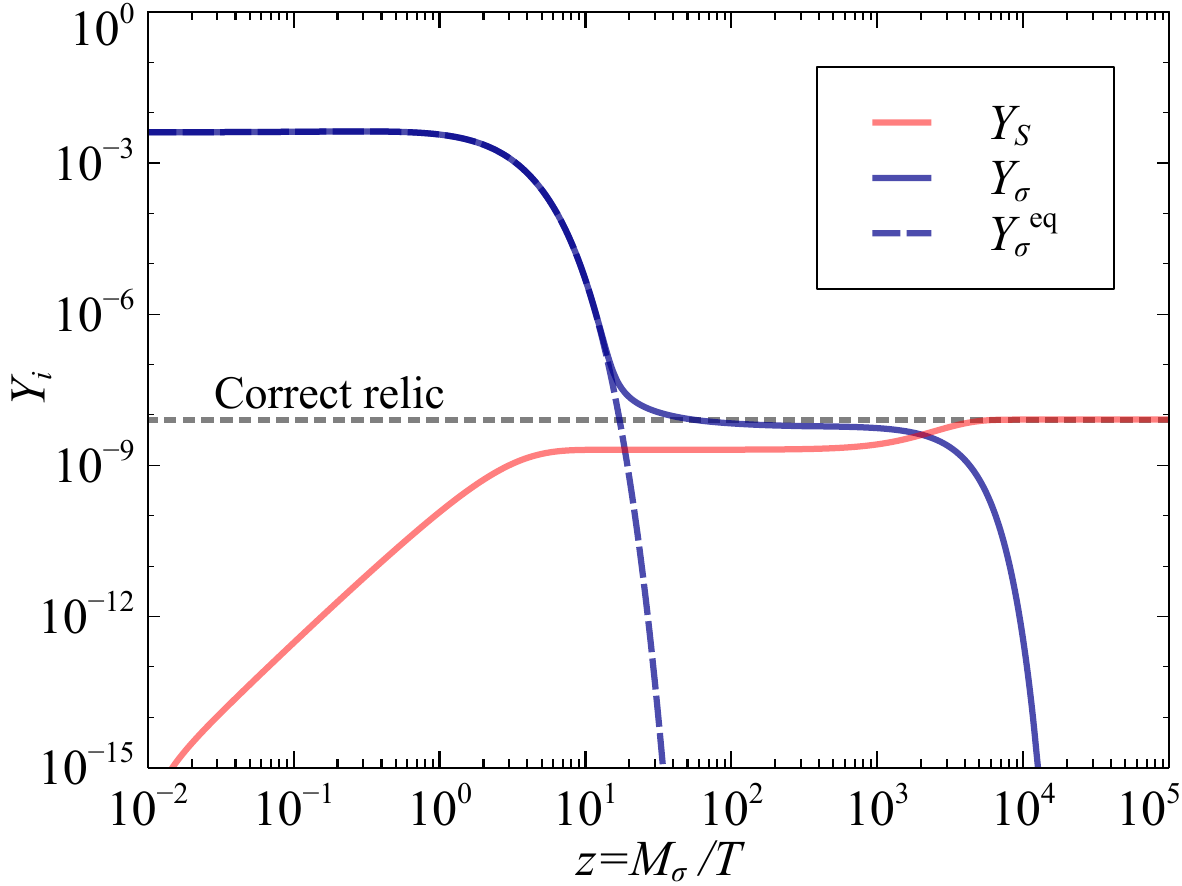}
\includegraphics[scale=0.4]{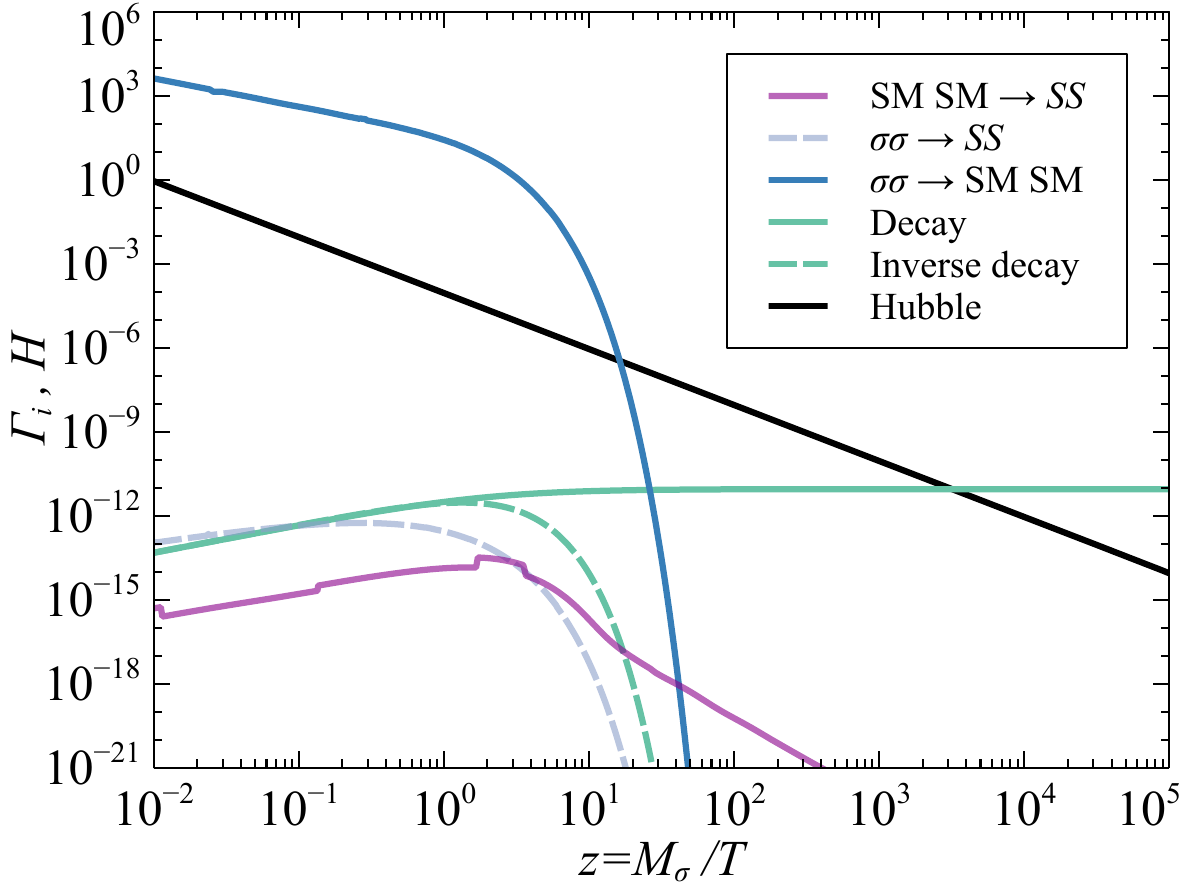}
\caption{\textit{Left:} Cosmological evolution of the DM, $S$ and the singlet scalar, $\sigma$ with respect to $z=M_\sigma/T$ for BP2 as mentioned in Table \ref{tab:tab1}. \textit{Right:} Comparison of the interaction rates of different processes is shown with different colors as mentioned in the inset of the figure.}
\label{fig:evolu2}
\end{figure*}
\begin{figure*}[h]
\centering
\includegraphics[scale=0.4]{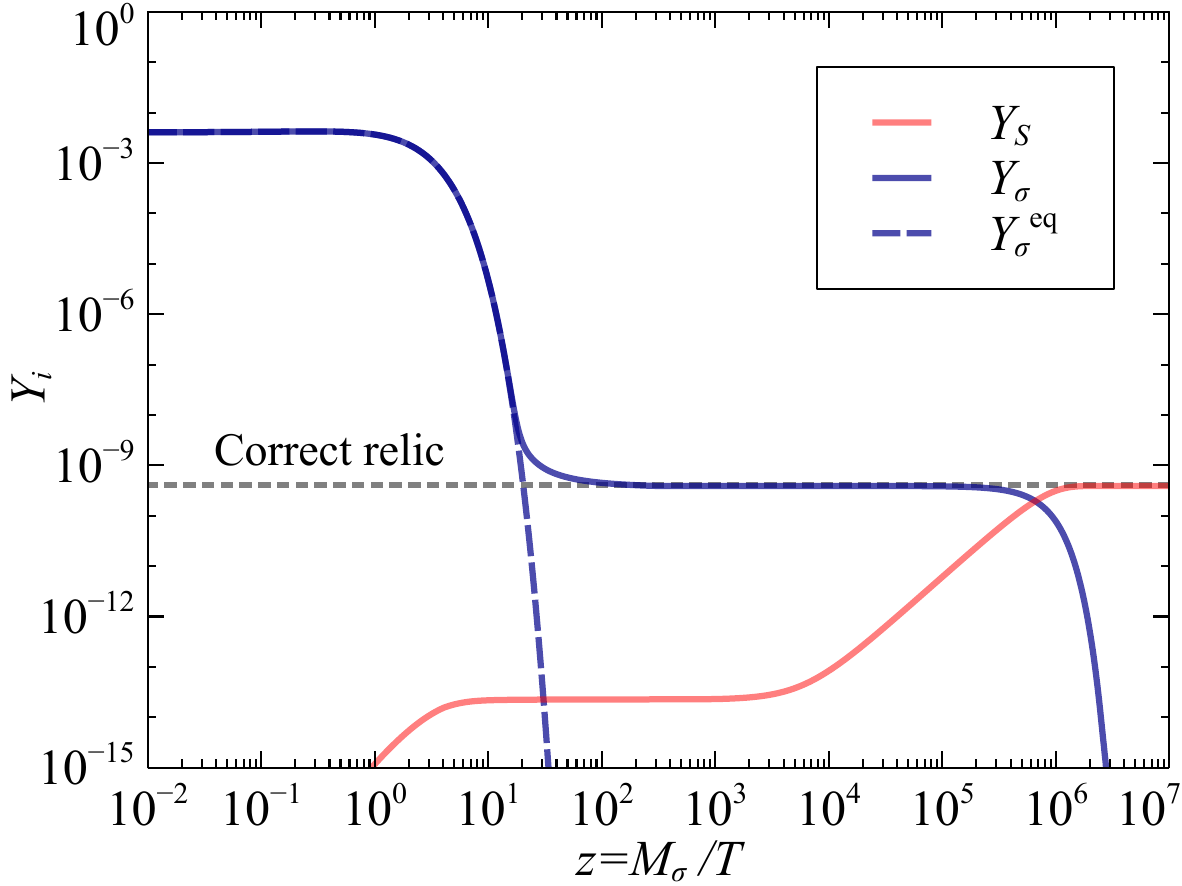}
\includegraphics[scale=0.4]{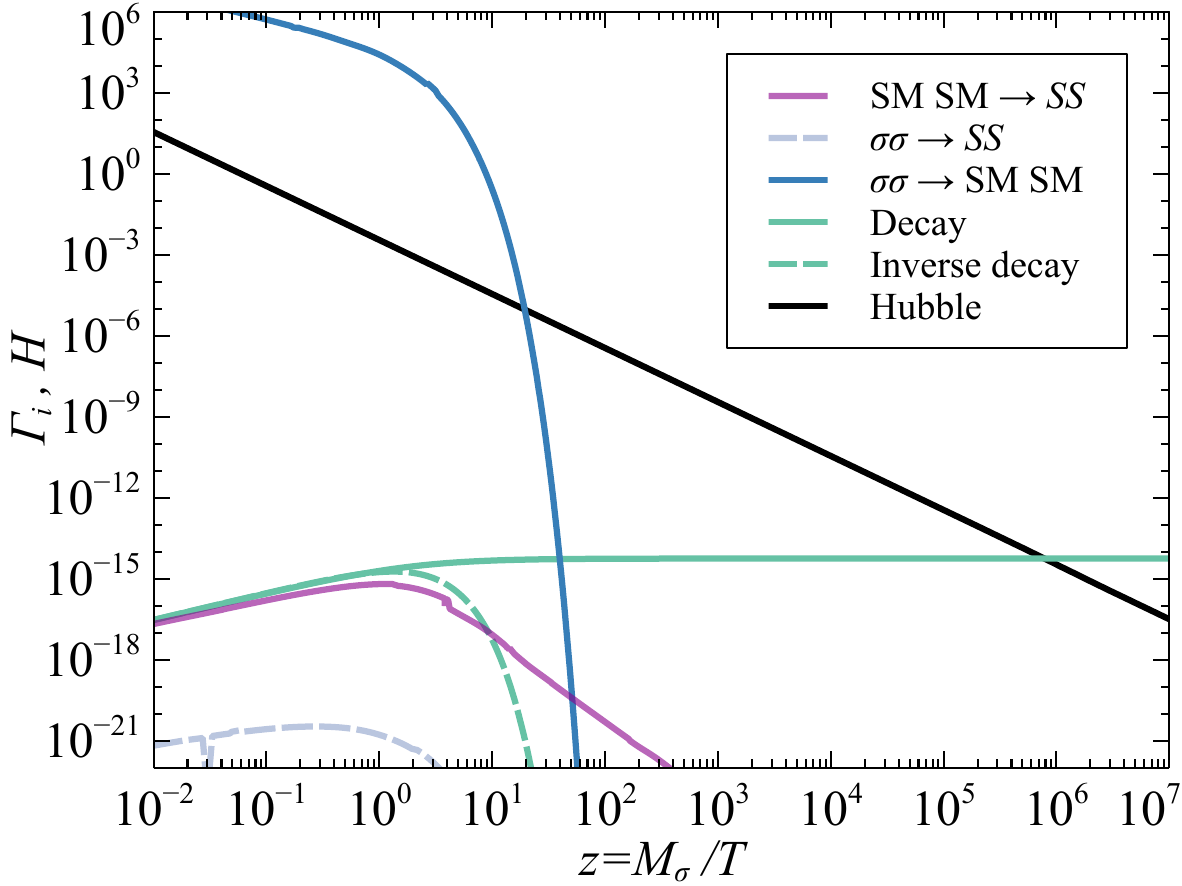}
\caption{\textit{Left:} Cosmological evolution of the DM, $S$ and the singlet scalar, $\sigma$ with respect to $z=M_\sigma/T$ for BP3 as mentioned in Table \ref{tab:tab1}. \textit{Right:} Comparison of the interaction rates of different processes is shown with different colors as mentioned in the inset of the figure.}
\label{fig:evolu3}
\end{figure*}
\begin{figure*}[h]
\centering
\includegraphics[scale=0.4]{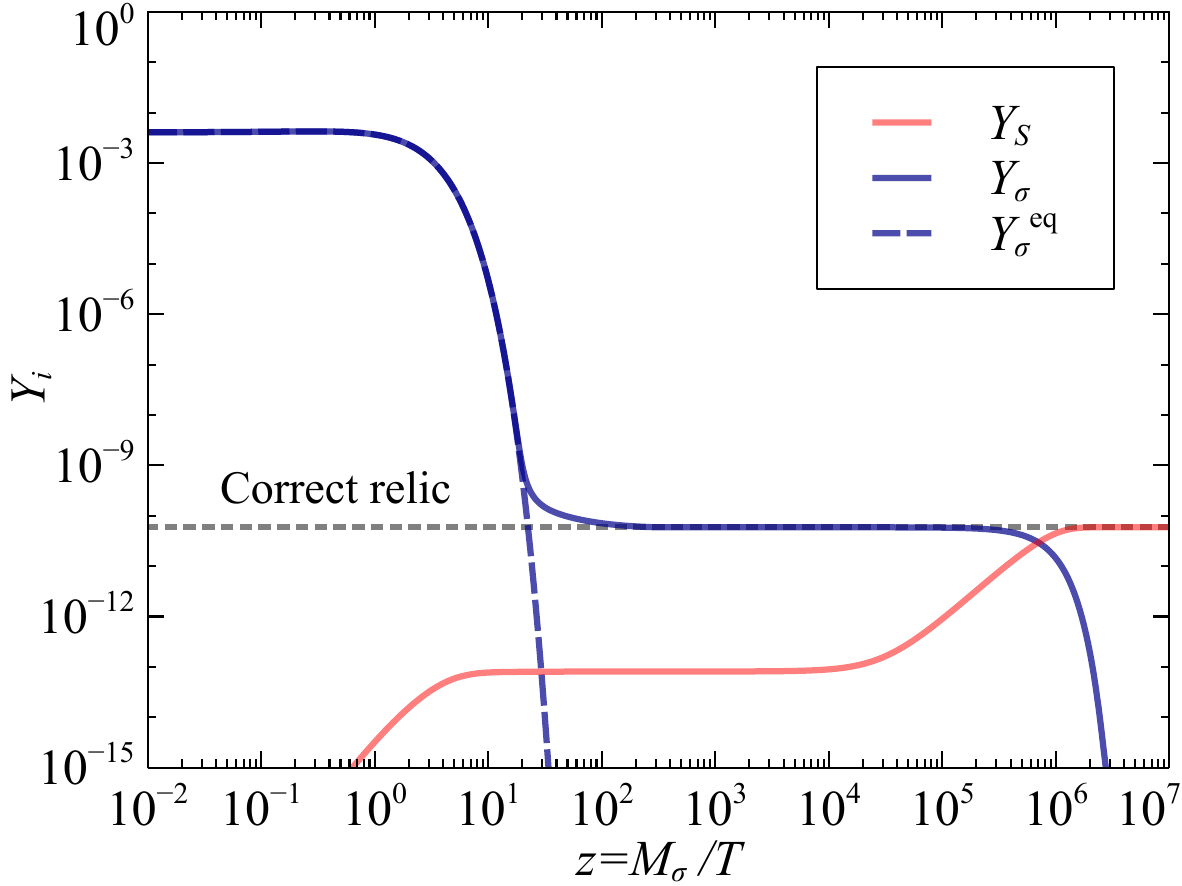}
\includegraphics[scale=0.4]{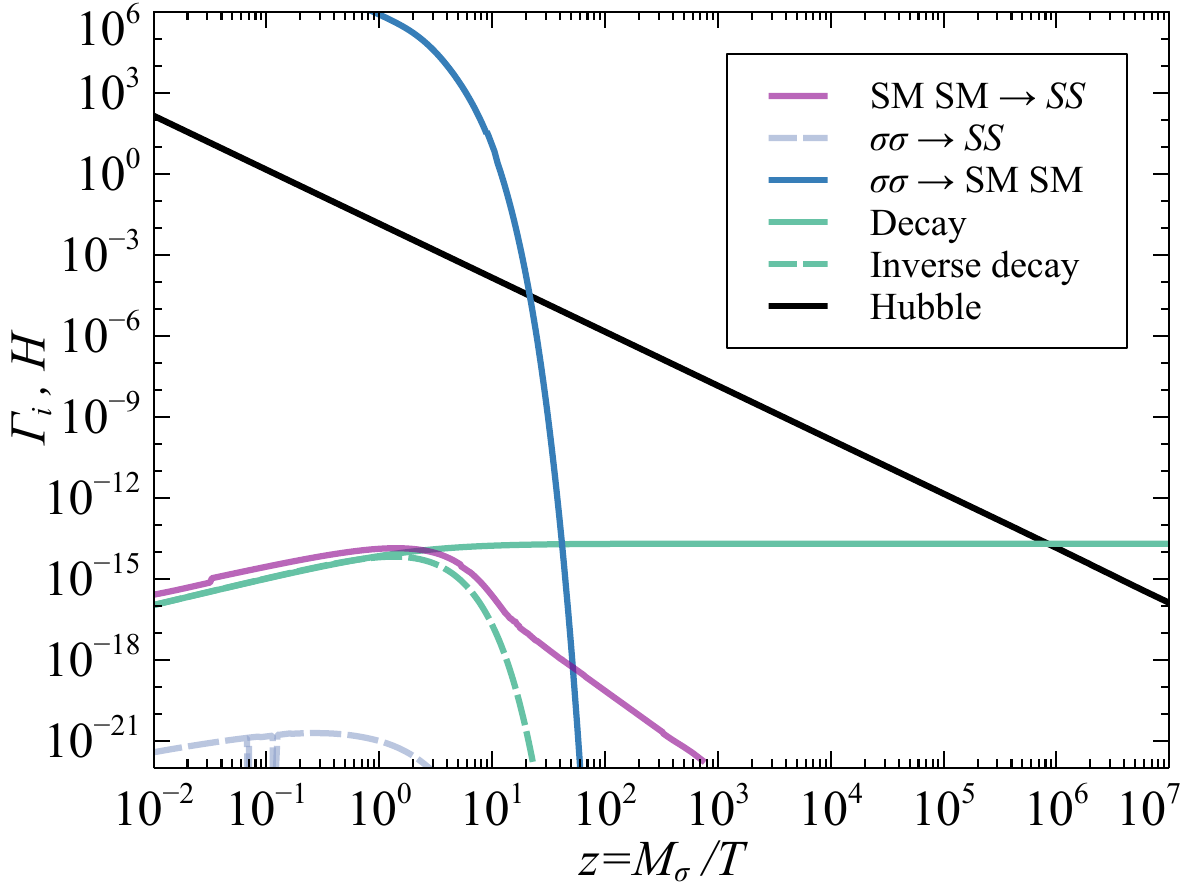}
\caption{\textit{Left:} Cosmological evolution of the DM, $S$ and the singlet scalar, $\sigma$ with respect to $z=M_\sigma/T$ for BP4 as mentioned in Table \ref{tab:tab1}. \textit{Right:} Comparison of the interaction rates of different processes is shown with different colors as mentioned in the inset of the figure.}
\label{fig:evolu4}
\end{figure*}
Here we show the evolution of the BP2, BP3, and BP4 from Table \ref{tab:tab1} along with the interaction rates of different processes. For the BP2, DM mass is 53 MeV and the Yukawa coupling is $7.5\times10^{-9}$. The evolution is shown in the \textit{left} panel of Fig. \ref{fig:evolu2}. In the \textit{right} panel of this figure, we see that the $\sigma\sigma\rightarrow$SM SM annihilation processes go below the Hubble rate around $z\sim17$. This decouples the $\sigma$ from the thermal plasma around $z\sim17$. As the $y_S\sin\theta$ is larger here compared to BP1, we see that the DM gets produced significantly from the $2\rightarrow2$ processes. After freezing out, the $\sigma$ decays and produces the DM, and the DM settles with the correct relic around $T\sim 1500$ GeV. In Fig. \ref{fig:evolu3}, we show the evolution of the abundances of the particles and the interaction rate comparison in the \textit{left} and \textit{right} panel, respectively, for the BP3. Here, the $2\rightarrow2$ production is minimal. DM gets produced from the decay, and it freezes in around $T\sim35$ GeV. The abundance evolution and interaction rate comparisons for the BP4 are shown in the \textit{left} and \textit{right} panel of Fig. \ref{fig:evolu4}. The DM mostly gets produced from the decay of the scalar, and it reaches its correct value around $T\sim67$ GeV.


%

\end{document}